\documentclass[useAMS,usenatbib]{mn2e}

\usepackage{epstopdf,graphicx,amsmath,amssymb,verbatim} %EJ added verbatim for comment environment
\title[Chemical consequences of accretion bursts I]{Astro \& cosmo-chemical consequences of accretion bursts I: the D/H ratio of water}
\author[Owen, J.E. \& Jacquet E.]{James E. Owen\thanks{E-mail: jowen@cita.utoronto.ca} and Emmanuel Jacquet\thanks{E-mail: ejacquet@cita.utoronto.ca}\\
Canadian Institute for Theoretical Astrophysics, 60 St. George Street, Toronto, M5S 3H8, Canada.}

\newcommand{\msunyr}{M$_\odot$~yr$^{-1}$~}
\newcommand{\msun}{M$_\odot$~}
\newcommand{\dhw}{(D/H)$_{\rm water}$ }
\newcommand{\dhnw}{(D/H)}
\newcommand{\dhnb}{D/H}

\begin{document}

\pagerange{\pageref{firstpage}--\pageref{lastpage}} \pubyear{2002}

\maketitle

\label{firstpage}

\begin{abstract}
The D/H ratio of water in protostellar systems is a result of both inheritance from the parent molecular cloud and isotopic exchange in the disc. A possibly widespread feature of disc evolution, ignored in previous studies, is accretion bursts (or FU Orionis outbursts), which may thermally process a large fraction of the water. One proposed underlying mechanism for FU Orionis outbursts relies on the presence of a magnetically dead zone. Here we examine the evolution of (D/H)$_{\rm water}$ in 1D simulations of a disc's evolution that include dead zones and infall from an envelope with given D/H ratio in the infalling water ($\sim 10^{-3}$),  and compare the results with similar calculations without dead zones.  We find that the accretion bursts result in a significantly lower (D/H)$_{\rm water}$ ratio and a more extended region (radius up to $\sim 1-3$ AU) where water is equilibrated with hydrogen gas (D/H=$2\times 10^{-5}$), when compared to burst-free models. Solar system constraints suggest that our solar nebula either experienced no accretion bursts and had a Schmidt number $\lesssim 0.2$ or had a Schmidt number closer to ``nominal'' values ($\sim 1$) and experienced several accretion bursts. Finally, future observations of (D/H)$_{\rm water}$ in protoplanetary discs will allow inferences about angular momentum properties of the disc during disc building and the role of accretion bursts.

% We study the evolution of the D/H ratio of water in a layered protoplanetary disc with a dead zone.  We find (D/H)$_{\rm water}$ is quickly equilibrated with the surrounding hydrogen gas above $\gtrsim 500$~K to a value of $\sim 2\times10^{-5}$, a temperature which the dead zone region ($\sim1-5$~AU) exceeds during an accretion burst.  Using 1D calculations of the disc's evolution that account for the chemical evolution of the D/H ratio of water in models with a steady infall rate, and in models which account for the formation and evolution of the disc. We find that the accretion bursts strongly influence the (D/H)$_{\rm water}$ profile and thermally process a large fraction of the disc's material. For a given D/H ratio (here $10^{-3}$) of water infalling from the parent molecular cloud, we find that accretion bursts result in an, on average, lower D/H ratio and a more distant radius where the D/H ratio reaches equilibration, compared to models without accretion bursts.  Comparing our simulations with solar-system constraints suggests that our solar nebula experienced no accretion bursts and had a Schmidt number $\lesssim 0.2$ or, more likely, experienced several accretion bursts and had a Schmidt number close to 1. Finally, future observations of the (D/H)$_{\rm water}$ ratio in protoplanetary discs can be used to make inferences about angular momentum properties of the disc during disc building and the role of accretion bursts.
\end{abstract}


\section{Introduction}

While most observed protoplanetary discs display accretion rates of order $10^{-8\pm 1}$~\msunyr \citep{hartmann98,ercolano14}, some objects, in particular FU Orionis systems \citep[e.g.][]{hartmann96}, have exhibited  episodes of enhanced accretion up to $\sim 10^{-4}$~\msunyr \citep{audard14} over observed time-scales of 10-100~years. It has been hypothesised that \textit{all} protostellar systems actually undergo such episodic accretion, equating to a small fraction (the ``duty cycle'') of their total lifetime, possibly concentrated in their early stages \citep{kenyon90}. Indeed this would alleviate the apparent discrepancy between the low observed accretion rates and that required to achieve the final mass of the protostars \citep{kenyon90,evans09,audard14}.  

  Although the mechanism underlying FU Orionis outbursts has not been established to date \citep{audard14}, a presently popular class of model uses as its only ingredients the two currently favoured drivers of transport in protoplanetary disks, namely the magnetorotational instability (MRI; \citealt{balbus98}) and the gravitational instability \citep{lodato04,lodato05,Durisenetal2007}. The ``gravo-magnetic cycle'', as it has come to be called \citep{martin11,martin13,martin14}, relies on the fact that a disc with a ``normal'' density is typically too cold ($\sim 100-300$~K) and too dense for its mid-plane to be sufficiently ionised to trigger the MRI, giving rise to a \textit{dead zone}. As the disc is gravitationally stable at such densities, accretion will only be possible in the surface layers where X-ray/UV (or cosmic ray) ionisation keeps the MRI alive \citep{gammie96}. However, a steady-state situation cannot be achieved and mass accumulates in the dead zone, until gravitational instability sets in, raising the disc's mid-plane temperature ($T_m$). Near $T_m\sim 800-1200$~K, thermal ionisation triggers the MRI throughout the entire thickness of the disc, resulting in an accretion burst which lasts $\sim 10^{3}-10^{4}$~years \citep[e.g.][]{armitage01,zhu10a,Ohtanietal2014,OA14}. Once the previously accumulated mass is dumped on the proto-star, the disc returns to the initial ``low state'' (or quiescent state). The cycle may then resume, with a duty cycle $\lesssim 1$~\% at low infall rates ($\sim 10^{-8}$~\msunyr) up to $\sim 30~\%$ at high infall rates ($\sim 10^{-6}$~\msunyr). 

  The significant changes in the disc's temperature and luminosity during such an accretion burst can chemically process a large fraction of the disc material. Several previous studies have shown the impact of accretion bursts on the chemical evolution of the envelope and disc. \citet{lee07,vorobyov13} studied the impact of accretion busts on the CO chemistry in the envelopes, in particular the large increase in luminosity, and found that repeated bursts could lead to repeated freeze-out cycles. These evolutionary CO cycles lead to possible observational evidence of accretion outbursts in young star forming cores \citep{kim11,kim12}. Additionally, \citet{visser12} used `single-point' models to study the impact of accretion bursts on the local evolution of various chemical species in the envelope, indicating that the outcome depends delicately on the balance between burst and chemical time-scales. However, what has not been considered in great detail is that this processed material can then be transported to smaller radii in the disc by advection and to larger ones by turbulent diffusion. These effects, which may persist for significantly longer than the duration of the bursts, may lend themselves to observable diagnostics; whether for extrasolar systems, e.g. with the increased resolution and sensitivity offered by {\it ALMA} and {\it Herschel}, or our own solar system, taking advantage of the rich body of cosmochemical evidence. To this end, in this series we set out to investigate the chemical consequences of accretion bursts.

  An astrochemically abundant species of choice in this regard is water \citep{water_science}, and in particular its \dhnb ~ratio \citep{VanDishoeketal2014,Ceccarellietal2014}. Observations of the interstellar medium and pre-stellar cores suggest that water starts deuterium-rich (\dhnb$\gtrsim 10^{-3}$; \citealt{Robertetal2006,Ceccarellietal2014})%, compared to the molecular hydrogen (\dhnb$\approx2\times10^{-5}$)
. Yet bulk chondrites and comets (with a further \textit{in situ} measurement to come for comet Churyumov-Gerasimenko \citealt{Haessigetal2013}) in our solar system display lower \dhw ratios (around $1-3\times 10^{-4}$), although some micron-scale D-rich hotspots in clays of the Semarkona chondrite do reach \dhw$\gtrsim 2\times 10^{-3}$ \citep{Delouleetal1998,Piani2012,piani14}. This in turn is much higher than the protosolar value (essentially that of molecular hydrogen) of $2\times 10^{-5}$ \citep{GeissGloeckler2003}, and reaction kinetics prohibit the conversion of low \dhw ratios to higher values in the protoplanetary disc \citep{Drouartetal1999}. The current interpretation is that the solar system inherited its water from the parent molecular cloud \citep{visser09,visser11}, and that this water \textit{partially} exchanged its deuterium with the nebular gas \citep{yang13}. Coupling this deuterium exchange with mixing throughout the disc produced the range of intermediate \dhw values we now observe \citep{Drouartetal1999,Mousisetal2000,Hersantetal2001,jacquet13,yang13,Albertssonetal2014}. The bottom line is that the \dhnw~ratio of water is a sensitive tracer of infall, temperature and turbulent mixing conditions. 

  All previous studies of the \dhw ~ratio assumed a constant turbulence level (often parametrised in terms of the Shakura \& Sunyaev $\alpha$, \citealt{ShakuraSunyaev1973}), and thus could not reproduce the gravo-magnetic cycle, and had no accretion outbursts. Additionally,  only \citet{yang13} considered the effects of infall, during which most of the outbursts are expected to take place \citep{armitage01,zhu10b}, along with the subsequent disc evolution. \citet{jacquet13} presented steady-state analytical calculations, indicating that in such types of model, a high efficiency of turbulent diffusion - with a Schmidt number $Sc$ of 0.2 or smaller - until the snow line and beyond would be required to account for the smallness of the \dhw ratio of carbonaceous chondrites and the Earth relative to Oort cloud comets; but as yet the realism of such a small Schmidt number is unclear \citep{jacquet13}. However, as isotopic exchange is most efficient for temperatures above $\sim 400-800$~K \citep[the ``equilibration temperature"\footnote{Its value depends somewhat on disc parameters, in particular $\alpha$, or the opacity, hence the range quoted.},][]{yang13,jacquet13}, clearly, outbursts, which may reach temperatures in excess of 1000 K, could significantly impact the \dhw profile of protoplanetary disc and possibly account for the low \dhw ratios of carbonaceous chondrites.

  In this work, we examine the effect of accretion outbursts arising from the gravo-magnetic limit cycle on the evolution of the \dhnw ~ratio of water using numerical models presented in Section \ref{sec:1dmodel}. Since this process has not been studied before, to gain insight we perform numerical experiments where the disc experiences a steady infall rate for the entire simulation (Section \ref{sec:steady}). %In these calculations the disc reaches a steady repeating limit cycle between a burst state with high temperatures and an accretion rate of $\sim 10^{-4}$~\msunyr and a quiescent state with low temperatures and an accretion rate of $\sim10^{-8}$~\msunyr.
 Secondly, we perform more realistic evolutionary calculations which include a disc building phase for a few hundred-thousand years followed by a stage of isolated evolution for several million years, that are meant to represent the evolution of an actual protoplanetary disc, e.g. the solar nebula (Section \ref{sec:building}). The implications are then discussed in Section \ref{sec:discussion} and the conclusions summarized in Section \ref{sec:conclusion}.

\section{one-dimensional disc model}\label{sec:1dmodel}
In order to investigate the effect of accretion bursts on the chemical evolution of the \dhw ratio we adopt a simple one-layer approach \citep{OA14}. Our model is a modified version of that originally  presented by \citet{armitage01} to study the impact of dead-zones in protoplanetary discs and associated accretion bursts, where we additionally include radial transport of heat by radiation and turbulence \citep[see][]{OA14}. Here we extend the model to include the transport of water due to advection and turbulent diffusion as well as the chemical evolution of the \dhw ratio. 

%THIS REPEATS OUTLINE Since this is the first study to include accretion bursts and dead-zones into the disc models we consider two calculations. First, we consider models where there is a constant rate of infall ($\dot{M}_{\rm in}$) onto the disc, which either result in a steady disc profile or a steadily repeating limit-cycle. Such steady infall models have been used to study the properties and occurrence of accretion bursts and as such will provide a useful starting point to understand the impact of accretion burst on the \dhw ratio. Secondly, we consider a more realistic evolutionary scenario where we build the disc and star from an infalling rotating core and then evolve the full star, disc \& core system similar to the previous study by \citet{yang13}.  

\subsection{Gas component}
The evolution of the surface density $\Sigma^g$ for a disc around a forming star, with a dead-zone is given by \citep[e.g.][]{armitage01}:
\begin{equation}
\frac{\partial\Sigma^g}{\partial t}=\frac{3}{R}\frac{\partial}{\partial R}\left\{R^{1/2}\frac{\partial}{\partial R}\left[\left(\nu_{\rm MRI}\Sigma^g_{\rm ac}+\nu_{\rm SG}\Sigma^g\right)R^{1/2}\right]\right\}+\dot{\Sigma}^g_{\rm in}\label{eqn:gas}
\end{equation}
where $\nu_i$ is the kinematic viscosity due to the MRI or self-gravity which we detail in Section~\ref{sec:viscosity} and $\dot{\Sigma}^g_{\rm in}$ is the rate of material falling onto the disc. $\Sigma^g_{\rm ac}$ is the surface density that is `MRI active' and is given by:
\begin{equation}
\Sigma^g_{\rm ac}=
\begin{cases} \Sigma^g & \textrm{if\,\,} T_m > T_{\rm crit} \textrm{\,\,or\,\,} \Sigma^g < 2\Sigma_{\rm layer} \\
2\Sigma_{\rm layer} & \textrm{otherwise}
\end{cases}
\end{equation}
For all our models where we include a dead-zone (and hence accretion bursts) we choose $T_{\rm crit}=800\,$K and $\Sigma_{\rm layer}=10^{2}\,$g~cm$^{-2}$. In simulations where we do not include a dead-zone (and thus no accretion bursts) we set $\Sigma^g_{\rm ac}=\Sigma^g$.

The evolution of the temperature at the mid-plane of the disc is given by \citep{OA14}:
\begin{eqnarray}
\frac{\partial T_m}{\partial t}&=&-v_R\frac{\partial T_m}{\partial R}+\frac{1}{\Sigma^g}\left[\frac{\Gamma}{C_p}-\frac{T_m}{\gamma\Theta R}\frac{\partial}{\partial R}\left(RF_\Theta\right)\right]\nonumber\\&&-\left(\frac{2H}{\Sigma^g RC_p}\right)\frac{\partial}{\partial R}\left(RF_R\right)\label{eqn:temp}
\end{eqnarray}
where $C_p$ is the heat capacity at constant pressure, $\gamma$ is the ratio of specific heats and $H=c_s/\Omega$ is the scale height of the disc, with $\Omega$ the Keplerian angular velocity and $c_s$ the isothermal sound speed. $\Theta$ is the `potential temperature' of the disc's mid-plane and following \citet{OA14} we define it as:
\begin{equation}
\Theta\propto T_m\left(\Sigma^g\Omega\right)^{\frac{2(1-\gamma)}{1+\gamma}}%=T_m\left[\frac{\Sigma}{\Sigma_{\rm MMSN}}\left(\frac{R}{1\,{\rm AU}}\right)^{-3/2}\right]^{\frac{2(1-\gamma)}{1+\gamma}}
,
\end{equation}
noting that Equation~\ref{eqn:temp} is independent of the choice of scaling for the potential temperature.The net heating rate ($\Gamma$) is given by:
\begin{equation}
\Gamma=\frac{9}{4}\nu\Sigma^g_{\rm ac}\Omega^2-\frac{16}{3\tau}\sigma_b T_m^{4},\label{eqn:gamma}
\end{equation}
where $\sigma_b$ is the Stefan-Boltzmann constant and $\tau=\Sigma^g\kappa/2$ is the vertical optical depth, with $\kappa$ the opacity. The two fluxes $F_{\Theta}$ \& $F_R$ are the radial heat fluxes due to turbulence and radiation respectively; which are given by:
\begin{equation}
F_\Theta=-\frac{3\nu\Sigma^g}{Pr}\frac{\partial \Theta}{\partial R},
\end{equation}
where $Pr$ is the Prandtl number (in most calculations we take $Pr=10$, the preferred value suggested by \citealt{OA14}), and
\begin{equation}
F_R=-\frac{4acT_m^3}{\kappa\rho}\frac{\partial T_m}{\partial R},
\end{equation}
with $a$ the radiation constant,  $c$ the speed of light and $\rho$ the mid-plane density.  For our model we adopt the \citet{belllin97} opacities and Equation~\ref{eqn:temp} is integrated as described in \citet{OA14}.

Our heating term (Equation~\ref{eqn:gamma}) does not account for stellar irradiation, which is known to dominate the heating in the outer disc \citep[e.g.][]{kenyon87,CG97,dalessio01}. Therefore, we do not allow the disc's temperature to drop below what would be obtained for a passively heated disc given by:
\begin{equation}
T_p=T_{1{\rm AU}}\left(\frac{R}{1\mbox{ AU}}\right)^{-1/2}
\end{equation}
We set $T_{1{\rm AU}}$ to:
\begin{equation}
T_{1{\rm AU}}=200\,{\rm K}\left[\frac{8}{9}\left(\frac{M_*}{1\,\mbox{M}_\odot}\right)+\frac{1}{9}\right]^{1/4}
\end{equation}
where the mass scaling is identical to that adopted by \citet{zhu10a}, based on a fit to a T Tauri star's birth-line from \citet{kenyon95}. Finally, the temperature is also restricted to be no less than 10~K at very large radius, to account for heating from the surrounding star-forming region.  In all calculations we assume an ideal equation of state with $\gamma=7/5$ and a mean molecular weight ($\mu$) of 2.35 amu as the majority of the disc material we are interested in remains molecular and neutral. 

\subsubsection{Viscosity and Self-Gravity}\label{sec:viscosity}
We treat the viscosity using an `alpha' model such that:
\begin{equation}
\nu_i=\alpha_i c_s H.
\end{equation}
%where $c_s$ is the isothermal sound speed. 
For MRI turbulence we set $\alpha_{MRI}=0.01$ and for angular momentum transport due to self-gravity we set $\alpha_{\rm SG}$ to
\begin{equation}
\alpha_{\rm SG}=\begin{cases} 0.01\left(\frac{Q^2_{\rm crit}}{Q^2}-1\right) & \mbox{if $Q\le Q_{\rm crit}$} \\
0 & \mbox{if $Q> Q_{\rm crit}$} \end{cases}\label{eqn:alpha_SG}
\end{equation}
where $Q$ is the Toomre parameter \citep{toomre64} given by:
\begin{equation}
Q=\frac{c_s\Omega}{\pi G \Sigma^g}
\end{equation}
The form of $\alpha_{SG}$ is not important and is merely designed to keep $Q\sim Q_{\rm crit}=2$ \citep[e.g.][]{lin_pringle87,lin_pringle90,zhu10a} as suggested by simulations of self-gravitating discs \citep[e.g.][]{lodato04,lodato05}.

\subsection{Water component}
We treat the water component of the disc as a passive scalar species, such that it is advected along with the gas and can diffuse with respect to the gas due to turbulence. Thus, the evolution of the water component is given by \citep[e.g.][]{jacquet13,yang13}:
\begin{eqnarray}
&&\frac{\partial \Sigma_i}{\partial t}=\frac{1}{R}\frac{\partial}{\partial R}\bigg[\frac{R}{Sc} \left(\nu_{\rm MRI}\Sigma^g_{\rm ac}+\nu_{SG}\Sigma^g\right)\frac{\partial}{\partial R}\left(\frac{\Sigma_i}{\Sigma^g}\right)\nonumber \\ && -R\Sigma_i v_R \bigg]+\dot{\Sigma_i}\label{eqn:s_water}
\end{eqnarray}
where we evolve Equation~\ref{eqn:s_water} for the two water species H$_2$O and HDO. $\dot{\Sigma}_i$ represents both infall onto the disc and the chemical evolution of the two species. Since the purpose of this work is to understand whether accretion bursts affect the evolution of the \dhw ratio rather than make detailed predictions, we do not attempt to model a full chemical network but adopt the simple model proposed by \citet{jacquet13} based on \citet{LecluseRobert1994}, where the \dhw ratio in water evolves as:
\begin{equation}
\frac{\rm d}{{\rm d} t}\left(\frac{D}{H}\right)=k_-(T)n_g\left[\left(\frac{D}{H}\right)_{\rm eq}(T)-\left(\frac{D}{H}\right)\right]\label{eqn:water_chem}
\end{equation}
where $n_g$ is the gas number density at the disc's mid-plane, $k_-(T)$ is the rate constant for the deuteration of water given by \citep{LecluseRobert1994}
\begin{equation}
k_-(T)=2\times10^{-22}\,\mbox{ cm$^{3}$ s$^{-1}$}\,\exp\left[-\left(\frac{5170\mbox{ K}}{T}\right)\right] %aren't you ashamed to make me add CGS units? NEVER -- That's typical British isolationism ---, pretty sure it was forced on us by the americans --- all the same! Whose fault is it that they also use the Imperial system?
\end{equation}
and (D/H)$_{\rm eq}$ is the equilibrium D/H ratio at a given temperature, which we take from \citet{richet77} assuming the D/H ratio in the background molecular hydrogen dominated disc is $2\times10^{-5}$. The latter values were found by \citet{yang13} to agree reasonably well with the values obtained from their more detailed chemical network. In all calculations we assume, like \citet{yang13}, that the \dhw ratio of the in-falling primordial water is $10^{-3}$. 

\subsection{Infall Models}\label{sec:infall}
The infall profile adopted depends on the evolutionary model (steady infall vs. disc building calculations). For the steady infall problem we use a Gaussian infall profile which peaks at a radius $R_{\rm in}$ and has standard deviation $0.1R_{\rm in}$ \citep[e.g.][]{OA14}.  For the full evolutionary calculations, where we account for disc building from a uniformly rotating collapsing core, we follow \citet{dullemond06_db,yang12,yang13} and adopt the \citet{hueso05} infall profile:
\begin{equation}
\dot{\Sigma}^g=\frac{\dot{M}}{8\pi R_c(t)^2}\left(\frac{R}{R_c(t)}\right)\left[1-\left(\frac{R}{R_c(t)}\right)^{1/2}\right],
\end{equation} 
where the instantaneous centrifugal radius ($R_c$) of the infalling material is given by angular momentum balance:
\begin{equation}\label{Rc}
R_c(t)=1.3\,{\rm AU}\,\left(\frac{\Omega_c}{10^{-14}\,\mbox{s}^{-1}}\right)^2\left(\frac{T_c}{15\,\mbox{K}}\right)^{-4}\left(\frac{M(t)}{0.5\,\mbox{M}_\odot}\right)^3
\end{equation}
where $\Omega_c$ is the core's rotation rate (considered to be constant in space and time), $T_c$ is the temperature of the core and $M(t)$ is the mass of the star \& disc. Finally, we assume the core follows an isothermal inside-out collapse \citep[e.g.][]{shu77} and use a constant infall rate of
\begin{equation}
\dot{M}=0.975\frac{c_c^3}{G}
\end{equation}
until the core mass is fully depleted, where $c_c$ is the sound speed for gas at a temperature of $T_c=15\:\rm K$ in all calculations \citep[as in][]{yang13}.  

\subsection{Numerical Method}\label{sec:num}
Our numerical code is based on the method detailed in \citet{owen14,OA14} where advective fluxes are reconstructed using a second order method and a Van-Leer limiter. Equations~\ref{eqn:temp} \& \ref{eqn:gas} are integrated explicitly using a scheme that is second order in space and first order in time using a staggered mesh and finite volume operators. In order to evolve the transport and chemistry of the water content we make use of operator splitting:  First, equation~\ref{eqn:s_water} without the chemical evolution terms is integrated explicitly using a scheme that is second order in space and first order in time. Secondly the chemical evolution is evolved by integrating Equation~\ref{eqn:water_chem} using an implicit update, such that:
\begin{eqnarray}
&&\frac{(D/H)^{n+1}-(D/H)^n}{\Delta t}=\nonumber \\ &&\;\;\;\; \;\;\;(k_-n_g)^{n+1/2}\left[\left(\frac{D}{H}\right)^{n+1}_{\rm eq}-\left(\frac{D}{H}\right)^{n+1}\right]
\end{eqnarray}
such a scheme is stable for equation that are both stiff and non-stiff over a simulation domain and thus appropriate for our problem. The simulations are initialised with the gas surface density set to the numerical floor of $\Sigma^g=10^{-20}$g~cm$^{-2}$ and the temperature set to 10~K.  We use a non-uniform radial grid, using power-law spacing \citep{owen14}. At the inner boundary we use a zero-torque boundary condition for the gas surface density and a free outflow boundary for the water species. For our steady infall calculations  we use 200 cells spaced between 0.1 and 66.8 AU and at the outer boundary we use a zero-flux boundary condition for both the gas and water species. Whereas for the disc building calculations we use  400 cells spaced between 0.047~AU and 6685~AU, at the outer boundary we use a zero shear boundary condition for the gas and a free-outflow boundary condition for the water.

 In Appendix~A, we benchmark our simplified scheme against the results of the more complicated chemical network used by \citet{yang13} and find very good agreement. Therefore, our simple scheme is appropriate for analysing the impact of accretion bursts on the D/H ratio in water. 

\section{Steady Infall Calculations}\label{sec:steady}

In order to better understand the effects of accretion bursts we consider a setup with steady infall; where the star's mass does not grow with time and is fixed to $M_*=1$~\msun. Such calculations have been previously used to study the properties of the accretion bursts themselves \citep{armitage01,zhu09,OA14} as they reach a steadily repeating limit cycle.    Initially, only gas falls onto the disc and it reaches a steady-limit cycle after approximately 0.3-0.6~Myr. After 1~Myr we then allow both water and gas to fall onto the disc and evolve the calculation for a further 3~Myr. We perform large sets of calculations, where  in all sets we vary the infall rate from $6.3\times10^{-9}$ to $10^{-5}$~\msunyr and the infall radius from 0.32 to 32~AU. We run one set of calculations with no dead-zone and thus no accretion bursts. Additionally, we run five further sets which include accretion bursts: three have a Prandtl number of 10 (the theoretically preferred value of \citealt{OA14}) and vary the Schmidt number taking $Sc=0.3$, 1.0 \& 3.0; finally we run two sets with a  Schmidt number of 1.0 but Prandtl numbers of 1 \& 100. 

\subsection{Results}
\begin{figure}
\centering
\includegraphics[width=\columnwidth]{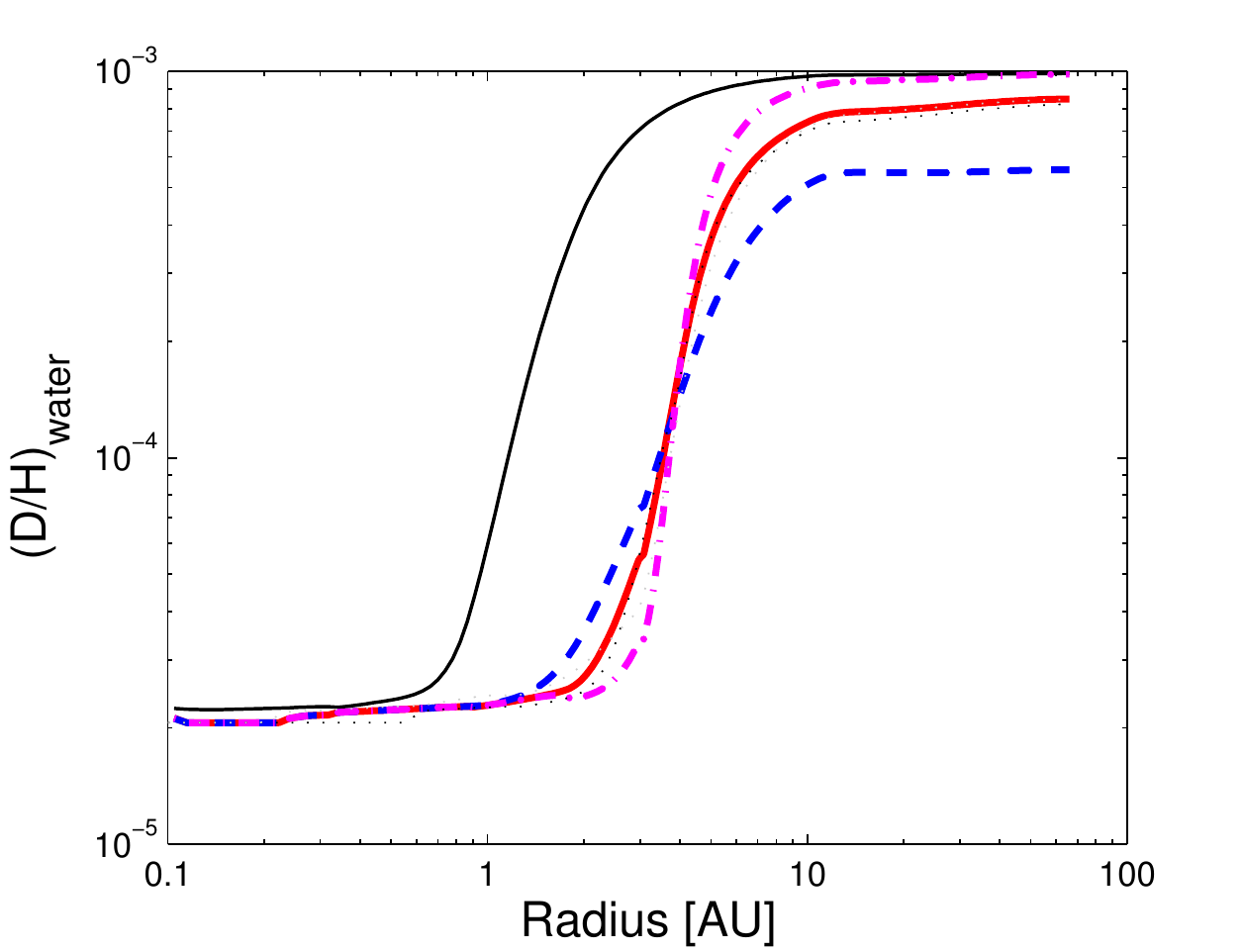}
\caption{The \dhw profile after 0.5~Myr of evolution in a disc with a steady infall rate of $\dot{M}=1.2\times10^{-6}$~\msunyr ~occurring at $R_{\rm in}=11.3$~AU. The thin solid line shows the result with no dead zone (and therefore no accretion bursts). The thick lines show simulations with accretion bursts with Schmidt numbers of (0.3 - dashed, 1.0 - solid \& 3.0 - dot-dashed). The thin dotted lines represent calculations with accretion bursts and a Schmidt number of 1.0 but Prandtl numbers of 1.0 (black) \& 100.0 (grey).}\label{fig:DH_profile1}
\end{figure}

In Figure~\ref{fig:DH_profile1}, we show the radial distribution of the D/H ratio in water 0.5 Myr after the water has been introduced into the disc (we plot the profile during quiescence as the D/H ratio shows a transient during the burst) for a simulation with $\dot{M}=1.2\times10^{-6}$~\msunyr and $R_{\rm in}=11.3$~AU in the six sets of calculations. 

These profiles clearly show that accretion bursts have a strong impact on the D/H ratio distribution.  With accretion bursts the radius at which the D/H ratio reaches the fully equilibrated value of $\sim 2\times10^{-5}$ (the ``equilibration radius''; where the equilibration time-scale is comparable to the transport time-scale c.f. \citealt{jacquet13}) increases from roughly 1~AU to 3~AU. Additionally, it decreases the D/H ratio at large radius in the disc from the unprocessed value ($10^{-3}$) with no accretion bursts to $\sim 5\times10^{-4}$ with accretion bursts and a Schmidt number of 0.3. Finally, we note that the value of the Prandtl number makes little difference. This is likely due to the fact that while the Prandtl number does affect the individual properties of an accretion burst (e.g. duration) it does not effect the duty-cycle of the burst significantly \citep{OA14} and hence the fraction of disc material processed by the accretion burst when averaged over a time-scale longer than the burst time-scale.

\begin{figure*}
\centering
\includegraphics[width=\textwidth]{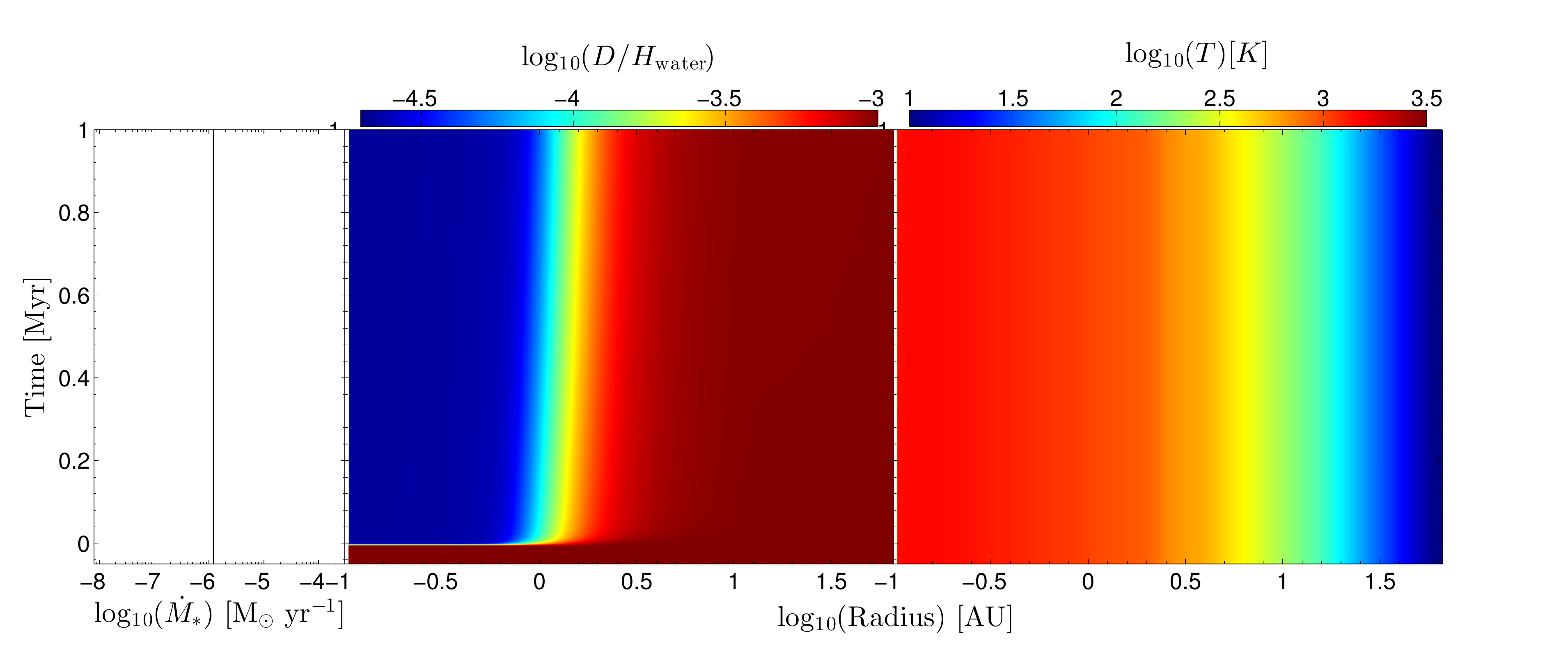}
\caption{Evolution of a steady infall calculation with $\dot{M}=1.2\times10^{-6}$~\msunyr and $R_{\rm in}=11.3$~AU with no dead zone and a Schmidt number of 1.0. The left-hand panel shows the accretion rate as a function of time. The middle panel is a space-time plot showing the evolution of the \dhw ratio and the right panel is a space-time plot for the temperature evolution. Zero-time has been reset to the point at which water is introduced into the system.}\label{fig:space_time_NB}
\end{figure*}

\begin{figure*}
\centering
\includegraphics[width=\textwidth]{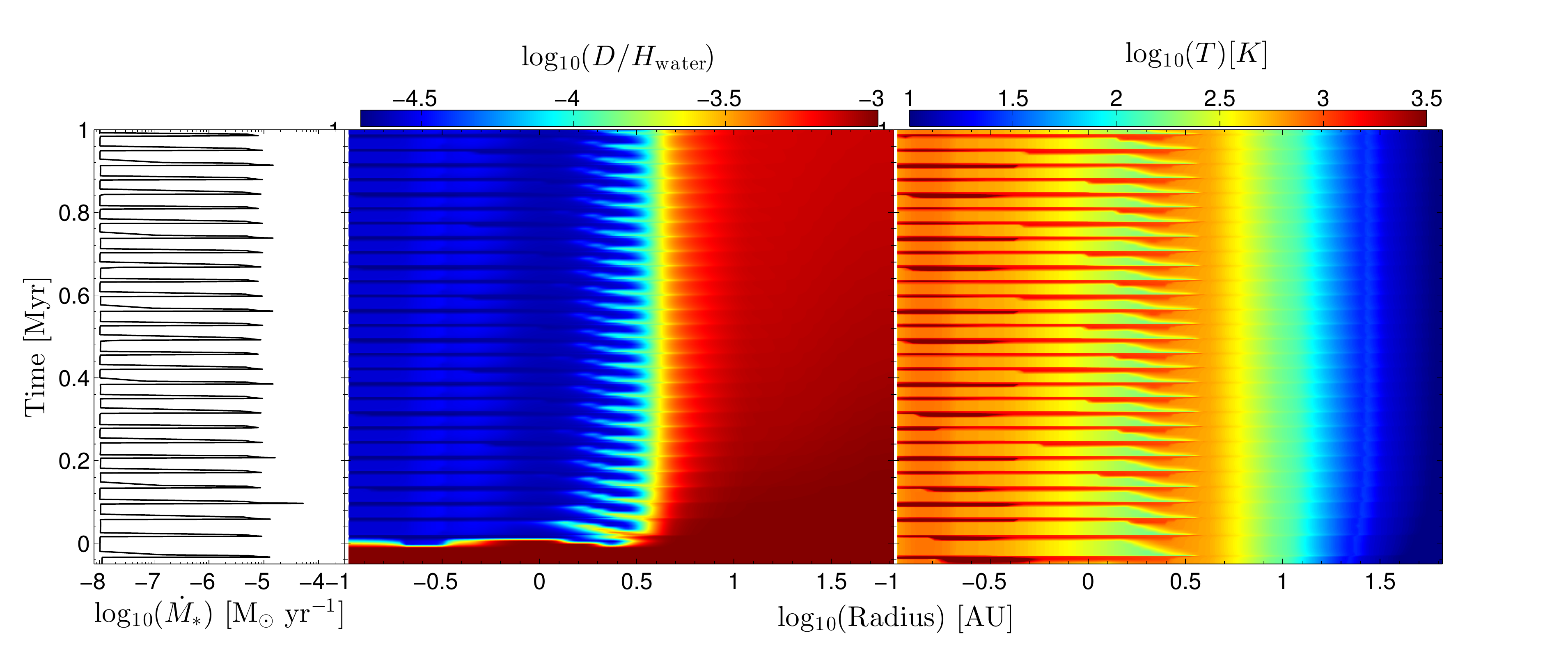}
\caption{Same as Figure \ref{fig:space_time_NB}, for a steady infall calculation with $\dot{M}=1.2\times10^{-6}$~\msunyr and $R_{\rm in}=11.3$~AU with accretion bursts and a Schmidt number of 1.0. %The left-hand panel shows the accretion rate as a function of time. The middle panel is a space-time plot showing the evolution of the $(D/H)_{\rm water}$ ratio and the right panel is a space-time plot for the temperature evolution. Zero-time has been reset to the point at which water is introduced into the system.
}\label{fig:space_time_B}
\end{figure*}

In Figures~\ref{fig:space_time_NB} \& \ref{fig:space_time_B} we show the space-time evolution of the temperature (right-panel) and \dhw (middle-panel), along with the accretion rate evolution (left-panel). 
Figure~\ref{fig:space_time_NB} shows that the \dhw ratio quickly reaches a steady state distribution where the equilibration radius occurs at 1~AU (the temperature here is essentially due to viscous dissipation), in agreement with the analytic models of \citet{jacquet13}. However, Figure~\ref{fig:space_time_B} shows that, in the presence of accretion bursts, the \dhw ratio never reaches a steady state; but like the accretion rate reaches a steadily repeating pattern, where incoming material is accreted into the dead zone region with a \dhw$\approx10^{-3}$, which is then equilibrated in the dead zone region when the disc goes through an accretion burst, which raises the temperature well above the equilibration temperature to $\sim 1000$~K. Thus, the dead zone is able to thermally process a larger fraction of the disc material, such that the equilibration radius is considerably further out than predicted by models that do not include accretion bursts. Furthermore, as well as being carried by advection along with the gas, this newly equilibrated water in the dead zone region can diffuse out to large radius ($\gtrsim 10$~AU) in the disc lowering the (time-independent) \dhw ratio. In the case with accretion bursts, this is more efficient than without because equilibration occurs at a larger radius.

\begin{figure*}
\centering
\includegraphics[width=\textwidth]{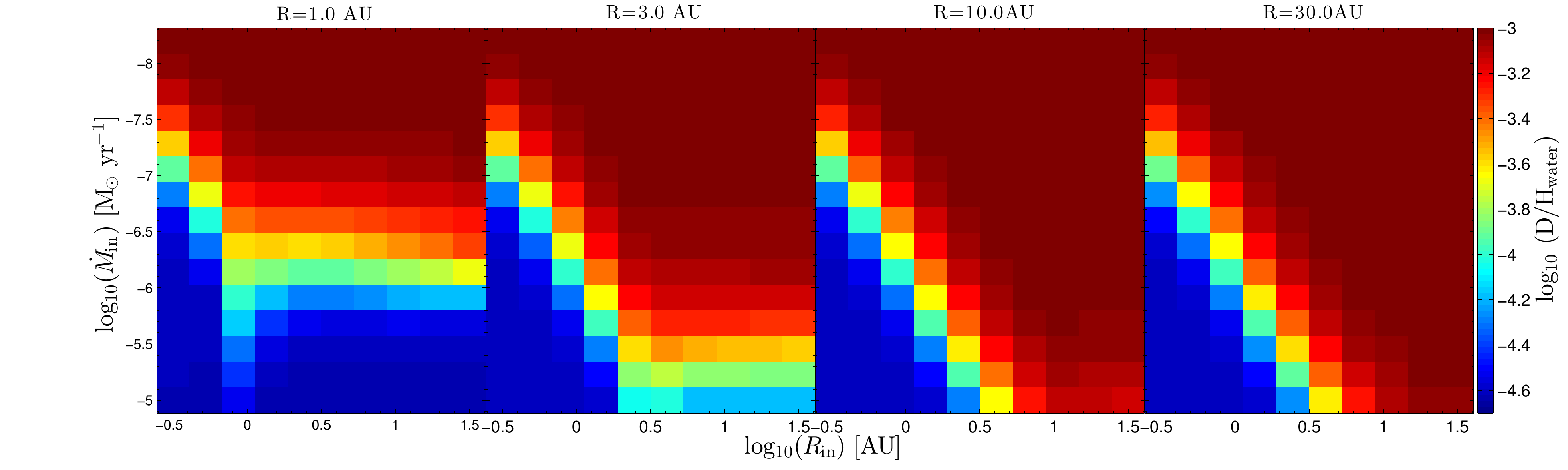}
\caption{Colour maps showing the \dhw ratio at 1.0, 3.0, 10.0 \& 30.0 AU in the protoplanetary disc with no dead-zone and a Schmidt number of 1.0 after of 0.5~Myrs of evolution.}\label{fig:panel_NB}
\end{figure*}
  
\begin{figure*}
\centering
\includegraphics[width=\textwidth]{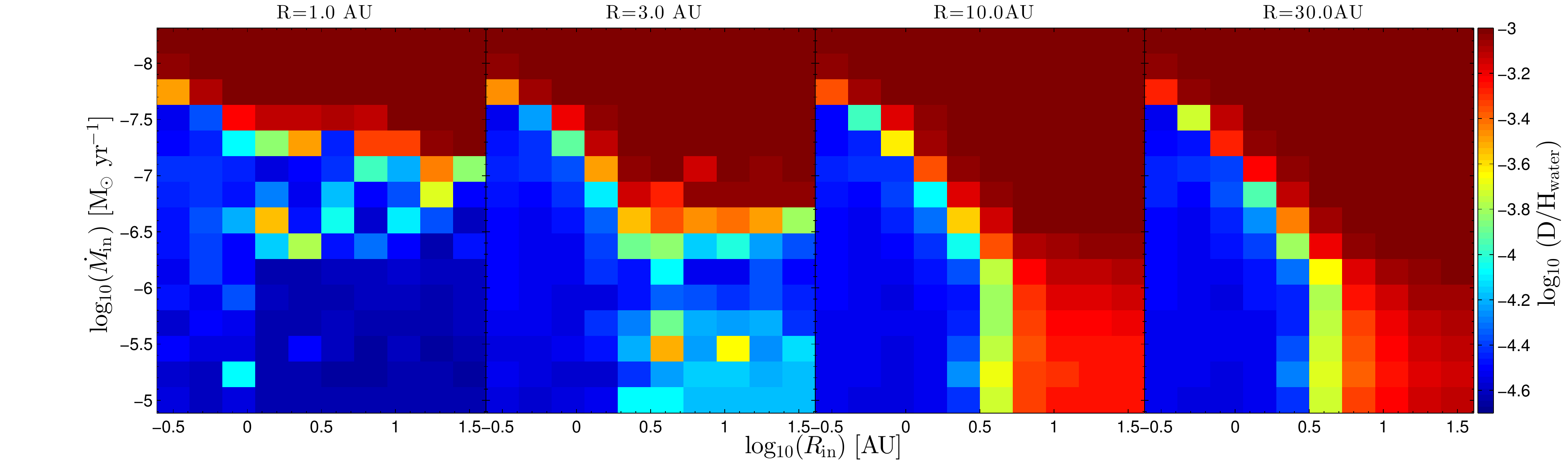}
\caption{Colour maps showing the \dhw ratio at 1.0, 3.0, 10.0 \& 30.0 AU in the protoplanetary disc with accretion bursts and a Schmidt number of 1.0 after 0.5~Myrs of evolution. The larger amplitude variations in the \dhw ratio in the predominately equilibrated regions is a measure of the temporal variability at a given location seen in Figure~\ref{fig:space_time_B} and discussed above.}\label{fig:panel_SC1}
\end{figure*}

\subsection{Parameter study}

In Figures~\ref{fig:panel_NB} and \ref{fig:panel_SC1} we show the \dhw ratio measured at different locations in the disc after 0.5 Myrs of evolution for the whole range of $\dot{M}$ and $R_{\rm in}$ examined. Figure~\ref{fig:panel_NB} shows the \dhw ratio at 1.0, 3.0, 10 \& 30 AU for the simulation set with a Schmidt number of 1.0 and no accretion bursts. This shows that at high accretion rates and small infall radii, most of the in-falling water is equilibrated before it can be transported to large radii resulting in a fully equilibrated disc. At low accretion rates $\lesssim 3\times10^{-8}$~\msunyr the disc is too cold to equilibrate the water except at very small radii ($\lesssim 1$~AU) and at intermediate values one gets a profile similar to that shown in Figure~\ref{fig:DH_profile1} where the water is equilibrated inside some radius, the equilibration radius, which is an increasing function of mass accretion rate. Essentially, the warmer the disc region onto which the infalling matter arrives, the more equilibrated the water of the disc as a whole will be.

The simulation set with accretion bursts and Schmidt number of 1.0 is shown in Figure~\ref{fig:panel_SC1} with the \dhw ratio measured at 1.0, 3.0, 10.0 \& 30.0 AU after 0.5 Myr of evolution. This shows a similar trend with infall rate and infall radius as previously; however, with accretion bursts it is much more pronounced. Similarly, the general trend of larger equilibration radius and lower \dhw ratio at large radius in the specific case discussed above holds for the majority of the parameter space. Comparing these results with those found in Figure~\ref{fig:panel_NB} indicates that for the majority of the parameter space of interest the terrestrial region will have equilibrated water. However, in order for the \dhw ratio to be reduced from primordial values at large radius requires infall rates $\gtrsim 3\times10^{-7}$~\msunyr, which is expected to occur at early times \citep[e.g.][]{hartmann98}. However, the reduced \dhw ratio at large radius does appear to sensitively depend on the infall radius, with $R_{\rm in}\lesssim 3$~AU resulting in fully equilibrated water at large radius: essentially all the incoming water is thermally processed in the dead-zone before it is transported to large radius. Whereas with $R_{\rm in}\gtrsim 3$~AU the \dhw ratio is only reduced from primordial values by a factor of $\sim 2$, where the equilibrated water at large radius comes from the turbulent diffusion of material processed in the dead-zone that has then be transported to larger radius.

\section{Disc Building and Evolution Calculations}\label{sec:building}

\begin{figure}
\centering
\includegraphics[width=\columnwidth]{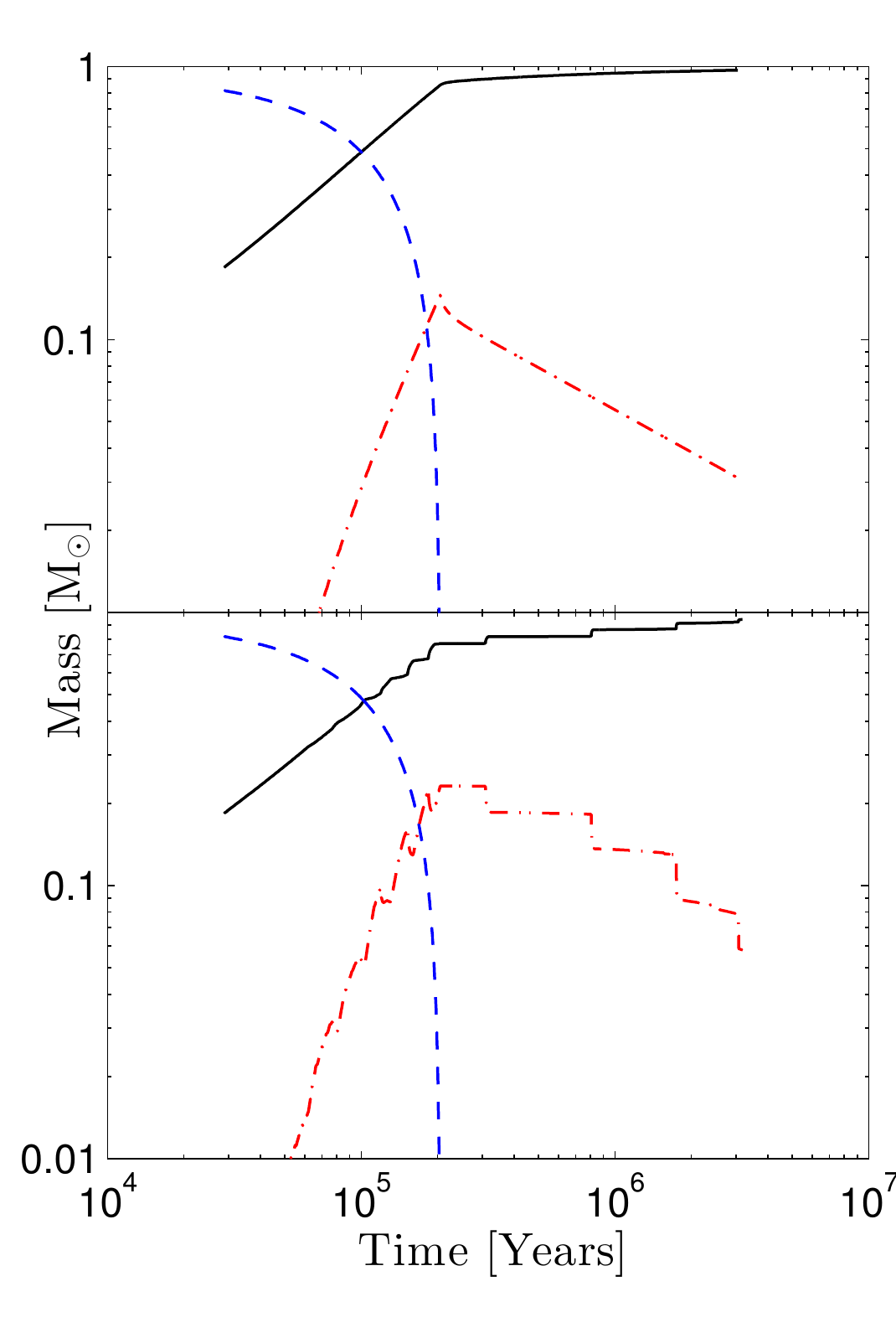}
\caption{The mass in the three components as a function of time: star (solid), cloud core (dashed) and disc (dot-dashed). The top panel shows a calculation with no accretion bursts and the bottom panel shows a calculation with accretion bursts. Both calculations have $\Omega_c=8.88\times10^{-15}$~s$^{-1}$ and a Schmidt number of unity. }\label{fig:mass_evolve}
\end{figure}

We now consider calculations where we build the disc from a uniformly rotating core. Initially the star has a mass of 0.2~\msun and the cloud core contains 0.8~\msun. %The cloud has a primordial \dhw ratio of $10^{-3}$. 
The cloud core then falls onto the disc as described in Section~\ref{sec:infall}. Any material that falls inside the inner boundary is automatically added to the star, whose mass is subject to an explicit update. Since we have seen above that the Prandtl number makes little difference to the results, we only perform calculations with $Pr=10$. We then perform a series of simulations where we vary the cloud's rotation rate $\Omega_c$ between $8\times10^{-16}$ \& $3.16\times10^{-13}$~s$^{-1}$, the Schmidt number between 0.1 \& 10 and whether or not the disc experiences accretion bursts.  

\subsection{Results}

In Figure~\ref{fig:mass_evolve}, we show how the individual components (star, cloud core \& disc) in our disc building calculations evolve during the simulation for a model with $\Omega_c=8.88\times10^{-15}$~s$^{-1}$. The top panel shows a model with no accretion burst and the bottom panel shows a simulation with accretion bursts. In this case infall takes approximately 0.2~Myr.

In the case with no accretion bursts the star rapidly approaches 1~\msun after infall ceases and the disc remains below 0.1~\msun for the majority of the evolution and steadily declines in a power-law fashion. However, in the case with accretion bursts, the dead zone traps material in the disc. This results in a more massive disc around 0.1-0.2~\msun for the majority of the evolution, which accretes material onto the star in rapid bursts. This evolution is similar to that found previously \citep[e.g.][]{armitage01,zhu10b,bae13} where discs that contain dead zones are generally more massive (and thus hotter) than discs that do not. 

\begin{figure*}
\centering
\includegraphics[width=\textwidth]{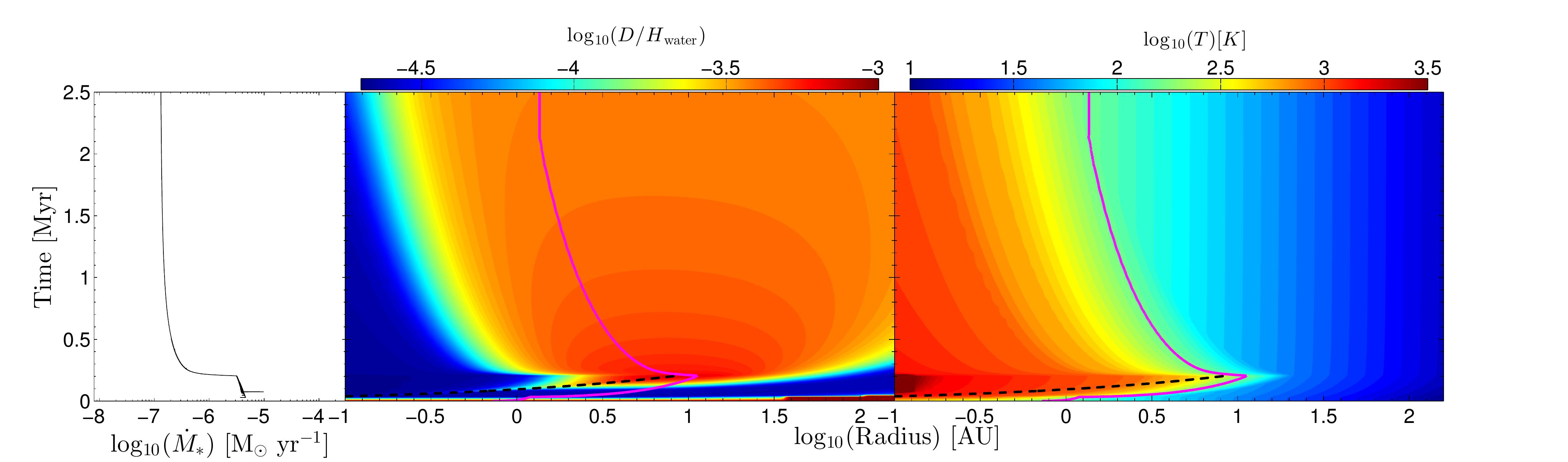}
\caption{Evolution of a disc building calculation calculation with $\Omega_c=8.88\times10^{-15}$~s$^{-1}$, no accretion bursts and a Schmidt number of 1.0. The left-hand panel shows the accretion rate as a function of time. The middle panel is a space-time plot showing the evolution of the $(D/H)_{\rm water}$ ratio and the right panel is a space-time plot for the temperature evolution. The thick solid line shows the time evolution of the water snow-line taken to be at 170~K. The mass evolution of the star, cloud core and disc are shown in the top panel of Figure~\ref{fig:mass_evolve}.}\label{fig:db_st_NB}
\end{figure*}

\begin{figure*}
\centering
\includegraphics[width=\textwidth]{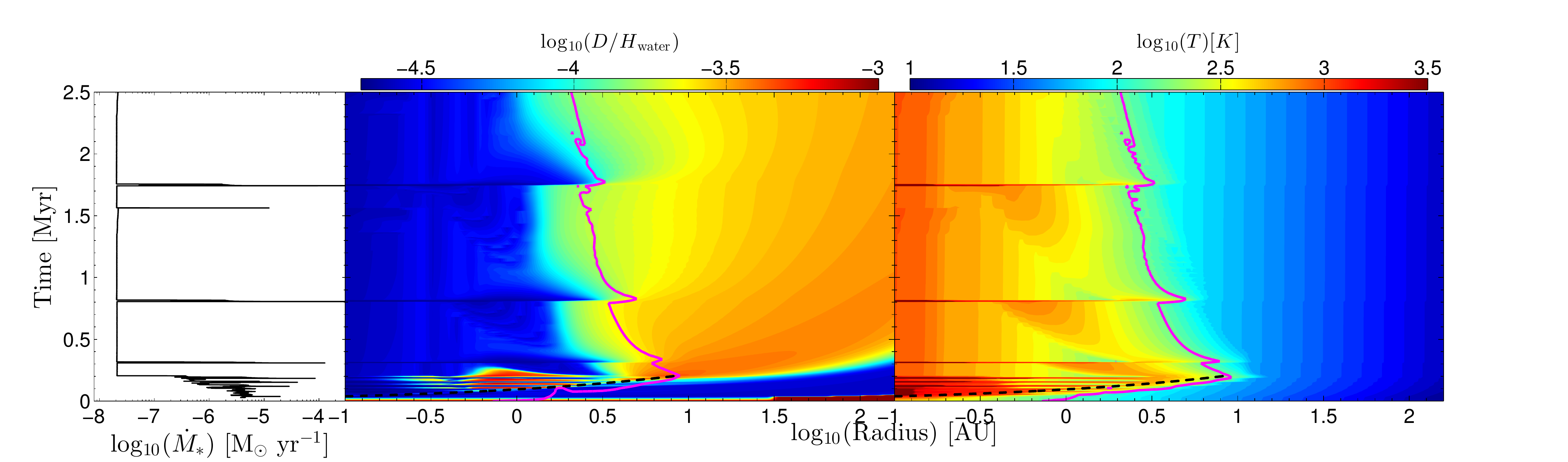}
\caption{Same as Figure \ref{fig:db_st_NB} for a disc building calculation calculation with $\Omega_c=8.88\times10^{-15}$~s$^{-1}$, accretion bursts and a Schmidt number of 1.0. %The left-hand panel shows the accretion rate as a function of time. The middle panel is a space-time plot showing the evolution of the $(D/H)_{\rm water}$ ratio and the right panel is a space-time plot for the temperature evolution.The thick solid line shows the time evolution of the water snow-line taken to be at 170~K. 
The mass evolution of the star, cloud core and disc are shown in the bottom panel of Figure~\ref{fig:mass_evolve}.}\label{fig:db_st_B}
\end{figure*}

In Figures~\ref{fig:db_st_NB} \& \ref{fig:db_st_B} we show the evolution of the accretion rate (left panel), centrifugal radius and snow-line radius taken at $T_M=170$~K (thick dashed line and thick solid line respectively in middle and right panels), \dhw ratio (middle panel) and disc temperature (right panel) as a function of time of the calculations shown in Figure~\ref{fig:mass_evolve}. Figure~\ref{fig:db_st_NB} shows the space-time evolution for a calculation with $\Omega_c=8.88\times10^{-15}$~s$^{-1}$ and no accretion burst. The accretion rate evolution shows that once the infall onto the disc ceases the accretion rate drops dramatically; this coincides with a large decrease in disc temperature, which then slowly declines with time. The middle panel shows that the \dhw ratio has a plateau around 10~AU and declines at both large and small radius. This evolutionary track was seen by \citet{yang13} and is also shown in our benchmark calculation discussed in the Appendix. At large radius this disc is primarily composed of material that fell at small radius into the hot disc and then was transported to large radius by decretion \citep[e.g.][]{dullemond06_cs,yang13}. The water that fell at early times into the hot inner disc was equilibrated and then transported to large radius resulting in a \dhw ratio close to the equilibrated value at large radius (although this radius increases with time). At small radius the disc is hot enough to rapidly equilibrate the water, although the equilibration radius moves inwards with time as the disc temperature falls as the disc's mass decreases. 

Figure~\ref{fig:db_st_B} shows the same calculation, but this time with accretion bursts. In this case, once infall ceases the accretion rate reaches $\sim 10^{-8}$~\msunyr, which then rapidly burst to $\sim 10^{-4}$ \msunyr occasionally. While the evolution of the temperature and \dhw ratio is similar at very large radius to the calculation without accretion bursts with a region of equilibrated water that is transported to large radius by decretion. However, we see that the presence of a dead-zone and accretion bursts dramatically changes the \dhw ratio profile in the inner regions of the disc. We also see an individual accretion burst has a transient effect that lasts $\sim$0.5~Myr.  Each accretion burst processes a large fraction of water in the disc, resulting in a lower peak \dhw ratio (as also seen in the steady infall calculations) and results in a equilibration radius that on average remains constant with time (at $\sim1$~AU) and does not move in with time like in the case without accretion bursts.

\begin{figure}
\centering
\includegraphics[width=\columnwidth]{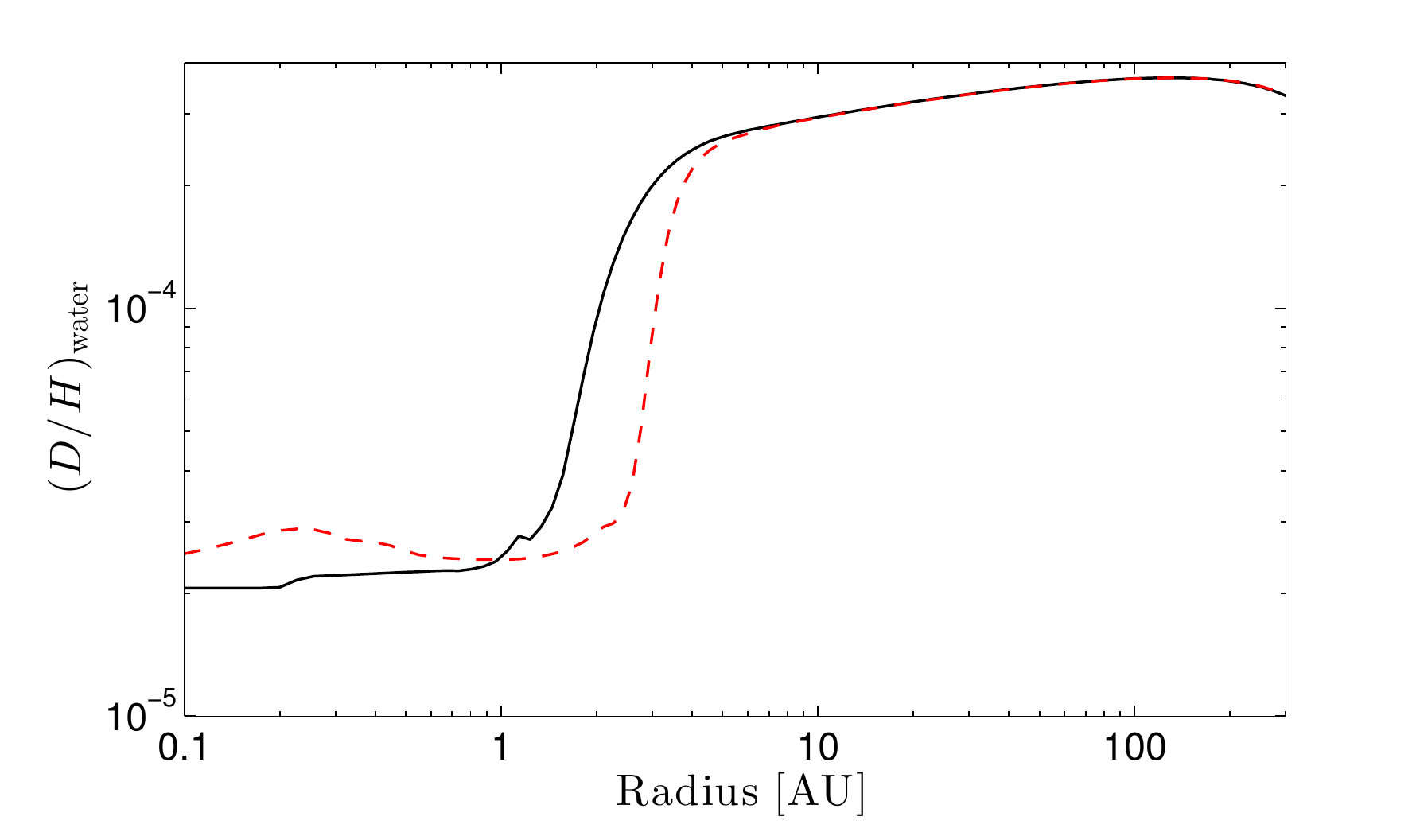}
\caption{The \dhw profile just before (0.795~Myr, solid) and after (0.819~Myr, dashed) an accretion burst. Shown for the simulation with $Sc=1.0$ and $\Omega_c=8.88\times10^{-15}$~s$^{-1}$ where the accretion burst occurs at $\sim$0.8~Myr (see Figure~\ref{fig:db_st_B}).}\label{fig:pre_post}
\end{figure}

The evolution of the \dhw profile during an accretion burst is shown in Figure~\ref{fig:pre_post}. 
where the solid line shows the \dhw profile just prior to an accretion burst and the dashed line shows the \dhw profile just after an accretion outburst. We see that the outburst equilibrates water in the dead zone ($\sim 0.5-3$~AU), decreasing the overall \dhw ratio of the entire disc. This newly equilibrated disc material is then transported by advection to smaller radius and to large radius by turbulent diffusion. This diffusion of newly equilibrated water to large radius results in a lower \dhw ratio at large radii in the disc, explaining lower peak \dhw ratios compared to burst-free models. 

\subsection{Parameter study}
\label{Parameter Evolution}

\begin{figure*}
\centering
\includegraphics[width=\textwidth]{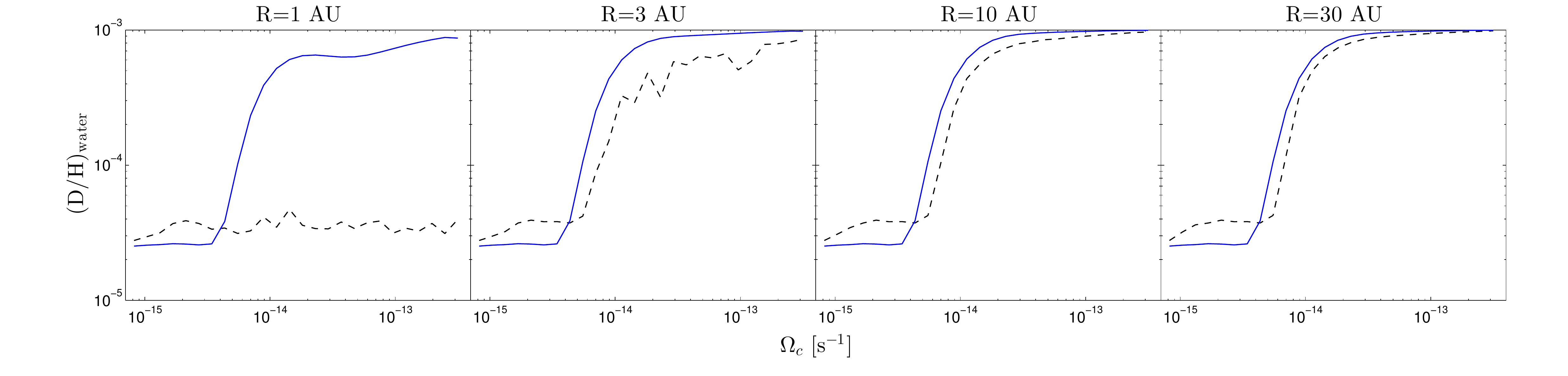}
\caption{The \dhw ratio as a function of the cloud core rotation rate ($\Omega_c$). The solid lines show calculations without accretion bursts and the dashed lines show calculations with accretion bursts, shown after 1~Myr of evolution. The four panels from left to right show disc radii of 1.0, 3.0, 10.0 \& 30.0 AU. All calculations used a Schmidt number of unity. The variation at small radius in the calculations with accretion bursts is a measure of the time-variability due to the variability due to the bursts.} \label{fig:DH_omega}
\end{figure*}

In Figure~\ref{fig:DH_omega}, we show how the core's rotation rate effects the \dhw profile. We plot the \dhw ratio for calculations with a Schmidt number of 1, both with (dashed) and without (solid) accretion bursts at 1, 3, 10 \& 30~AU after 1~Myr of evolution.  These show that there is a critical rotation rate below which the entire disc is equilibrated of $\sim 3\times10^{-15}$~s$^{-1}$ for the calculations without accretion bursts, and $\sim6\times10^{-15}$~s$^{-1}$ for the calculations with accretion bursts, corresponding to maximum centrifugal radii of $\sim 1$~AU and $\sim 4$~AU, respectively (see equation (\ref{Rc})). Below these critical rotation rates all infalling material lands in regions of the disc where the temperature is above the equilibration temperature. 

In the cases with accretion bursts at all rotation rates we find that at radii of $\sim 1$~AU the water is equilibrated, unlike the case without accretion burst where above the critical core rotation rate it returns to near primordial values. At large radii the trend seen from the steady infall calculations is also repeated here, where accretion bursts results in a lower \dhw ratio; however, the effect is less pronounced.  We note the variability seen in the 1.0 \& 3.0 AU is indicative of the variability in the \dhw ratio with time, in the case with accretion bursts. 

\begin{figure}
\centering
\includegraphics[width=\columnwidth]{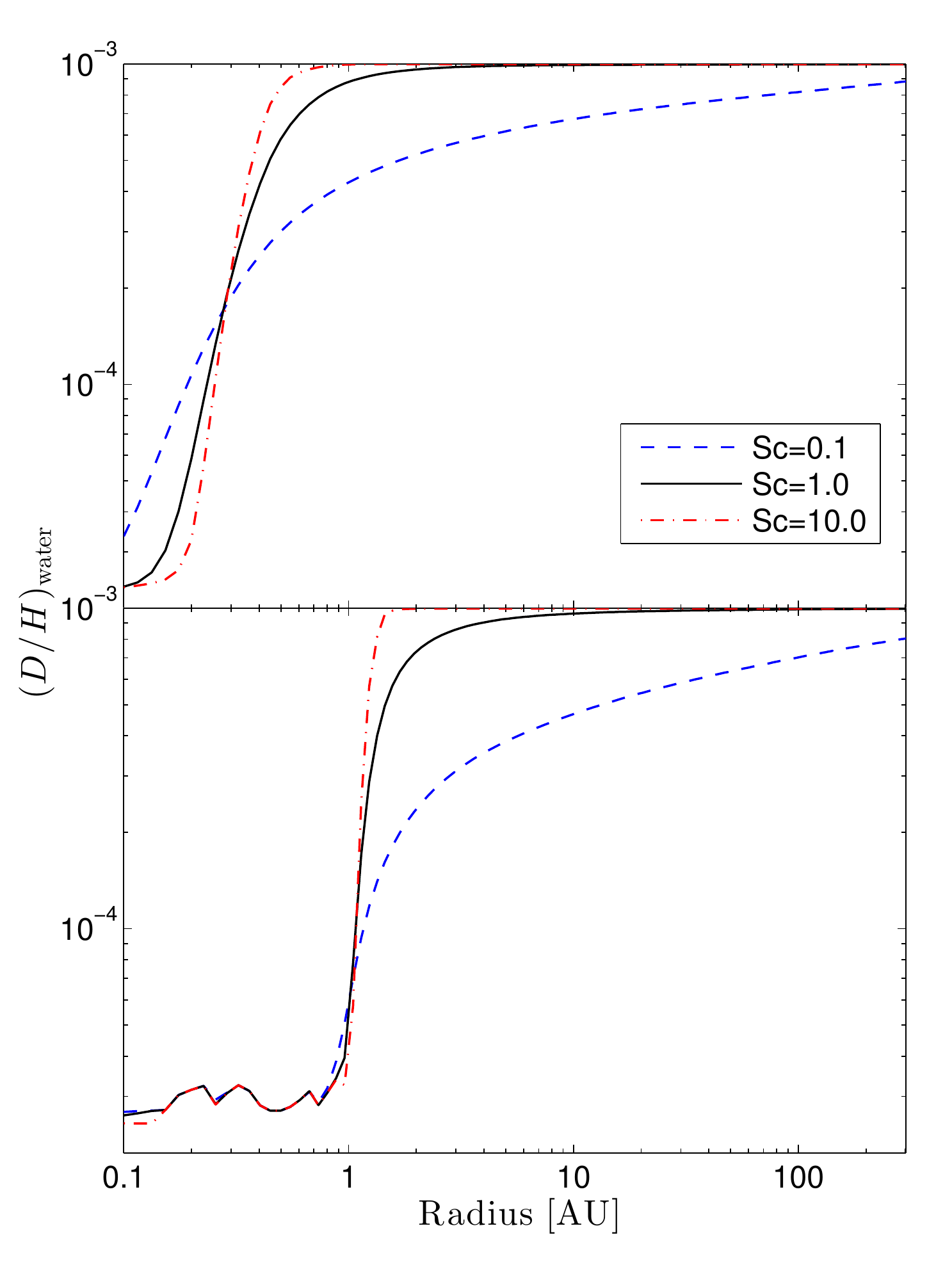}
\caption{\dhw ratio as a function of radius for calculations with $\Omega_c=3.16\times10^{-13}$~s$^{-1}$ for Schmidt numbers of 0.1 (dashed), 1.0 (solid) \& 10.0 (dot-dashed). The top panel shows calculations without accretion bursts the bottom panel shows calculations with accretion bursts.}\label{fig:profile_SC}
\end{figure}

Finally, in Figure~\ref{fig:profile_SC} we show the impact of the Schmidt number on the \dhw profile, where we plot our the \dhw ratio as a function of radius in our calculations with $\Omega_c=3.16\times10^{-12}$~s$^{-1}$ after 1~Myr of evolution with Schmidt numbers of 0.1 (dashed), 1.0 (solid) \& 10.0 (dot-dashed). The top panel shows the calculation without accretion bursts and the bottom panel shows the calculations with accretion bursts.  At small Schmidt numbers there is enhanced transport of equilibrated water to large radius resulting in a lower \dhw ratio at large radius (as seen by \citealt{jacquet13}). Conversely, larger Schmidt numbers result in sharper transitions from the equilibrated water %at small radius
 to the primordial value outward %at larger radius
 as expected from %due to 
the reduced efficiency of turbulent diffusion \citep[e.g.][]{jacquet13,owen14b}.

\section{Discussion}\label{sec:discussion}

%-General discussion of results: i) Bursts matter, ii) Bursts have a stronger influence of profile than Scmidt number (for sensible range). iii) Bursts lower the ratio at large radius and move the equilibrated radius to a slightly larger radius. 

Our calculations have shown that the presence of dead zones in protoplanetary discs, which can give rise to accretion bursts have a significant impact on the evolution of the \dhw ratio in water. This is not surprising since the \dhw ratio equilibrates at temperatures $> 400-800$~K \citep{jacquet13,yang13} which lies in-between the quiescent temperature of the dead-zone of $\sim 200$~K and the dead-zone temperature during an accretion burst of $\sim 1000$~K. Hence, un-equilibrated material that resides in the dead zone prior to an accretion burst is brought closer to equilibration by the higher disc temperatures.

%The evolution of the \dhw profile during an accretion burst is shown in Figure~\ref{fig:pre_post}. 
%\begin{figure}
%\centering
%\includegraphics[width=\columnwidth]{pre_post_burst}
%\caption{The \dhw profile just before (solid) and after (dashed) an accretion burst. Shown for the simulation with $Sc=1.0$ and $\Omega_c=8.88\times10^{-15}$~s$^{-1}$ where the accretion burst occurs at $\sim$0.8~Myr (see Figure~\ref{fig:db_st_B}).}\label{fig:pre_post}
%\end{figure}  
%where the solid line shows the \dhw profile just prior to an accretion burst and the dashed line shows the \dhw profile just after an accretion outburst. We see that the outburst equilibrates water in the dead-zone ($\sim 0.5-3$~AU), decreasing the overall \dhw ratio of the entire disc. This newly equilibrated disc material is then transported by advection to smaller radius and to large radius by turbulent diffusion. This diffusion of newly equilibrated water to large radius results in a lower \dhw ratio at large radii in the disc, explaining lower \dhw ratios at large radii in discs that experience accretion bursts.  

\subsection{Implications for astrophysical systems}

Our calculations have shown that the cloud core's rotation rate plays a strong role in setting the \dhw ratio at large radius in the disc. Thus, given the several order of magnitude spread in observed core rotation rates, with typical values $\sim 10^{-14}$ s$^{-1}$ \citep{goodman93,bg98,caselli02} then one would expect to observe a large spread in the \dhw values measured at large radii in protoplanetary discs. %Therefore, {\it Herschel} \citep[e.g.][]{visser13} or {\it ALMA} measurements of the \dhw ratio would allow inferences about the initial angular momentum of the disc. 
Therefore, Herschel \citep[e.g.][]{visser13}, ALMA or TMT measurements of the \dhw ratio would allow inferences about the initial angular momentum of the discs. Additionally, if the \dhw ratio could be spatially resolved then the properties of accretion bursts could be probed or spatial variations in the \dhw ratio then the
probe could be used to place constraints on the Schmidt
number, similar to the method detailed by \citet{owen14b}
for spatial variations at various molecular snow lines. Furthermore, combined observations of the crystalline silicate fraction \citep[e.g.][]{bouwman01} and \dhw ratio would allow the outward transport of crystalline materials by decretion during disc building hypothesis \citep[e.g.][]{dullemond06_cs} to be tested. Indeed discs with high crystallinity fractions at large radius should also possess lower \dhw ratios. 

%Furthermore, the high angular resolution ($\sim 0.01$ arcsec at 300~GHz) offered by {\it ALMA} opens up the possibility of spatially resolving the \dhw profile. At the closest Gould belt star forming regions (150~pc) this suggests a spatial resolution of $\sim 1.5$ ~AU (and possibly sub-AU in the closer disc TW Hya). In particular, such a resolution would be sufficient to resolve the equilibration radius in discs that experienced accretion bursts for all core rotation rates (see Figure~\ref{fig:DH_omega}). Whereas, our models suggest that if the majority of discs did not experience accretion bursts then {\it ALMA} would be unable to resolve the fall of in the \dhw ratio at small radii ($\lesssim 1$~AU). Furthermore, if {\it ALMA} were able to detect spatial variations in the \dhw ratio then the \textbf{steepness of the} profile could be used to place constraints on the Schmidt number \textbf{(see end of Section \ref{Parameter Evolution}), analogous} to the method detailed by \citet{owen14b} for spatial variations at various molecular snow lines. 

Thus, future observations of the \dhw ratio should be able to probe the angular momentum distribution of the disc during the its building and formation. Additionally, spatially resolved observations of the \dhw ratio should be able to determine whether the majority of protoplanetary discs experienced accretion bursts during their early evolution.     

\subsection{Solar system constraints}

%-Discussion in the context of solar system constraints. Future solar-system measurements
To first order, the D/H systematics in the solar system may be subsumed in two salient features \citep{Robertetal2006,Ceccarellietal2014}: 
\begin{itemize}
\item[(i)] Most Oort cloud comets show \dhw values of $\sim 3\times 10^{-4}$ (with some comets showing lower values), lower than expected in molecular clouds. 
\item[(ii)] Carbonaceous chondrites display D/H ratios significantly lower than cometary (around the terrestrial value $1.5\times 10^{-4}$). 
\end{itemize}
Both domains of variation are significantly higher than the protosolar value ($2\times 10^{-5}$) and significantly lower than the value observed in molecular clouds ($\gtrsim 10^{-3}$).

  In our framework, fact (i) indicates that infall has occurred on relatively small radii to allow significant equilibration before outward transport, which, as mentioned in the previous subsection, would also account for the presence in comets of crystalline silicates, also deemed to have originated in the inner solar system \citep{dullemond06_cs}. At the same time, the relatively large D/H ratio of cometary water compared to the protosolar value suggests that equilibration has not been complete. Judging from Figure \ref{fig:DH_omega}, that would at face value constrain the rotation rate to be within a factor of several of $10^{-14}$~s$^{-1}$ (with a factor of $\sim 2$ systematic difference between the outburst-free and outburst-laden cases). Of course, the infall model is very idealized, so the situation may not be as fine-tuned as it may seem (whether with or without outbursts, incidentally). It is also possible that comets, whose \dhw is somewhat variable, do not mark the \textit{peak} \dhw value of the disc, and whatever that may be, this latter value may still be consistent with the lower end of those determined for prestellar cores \citep[e.g.][]{Ceccarellietal2014}.

\begin{figure}
\centering
\includegraphics[width=\columnwidth]{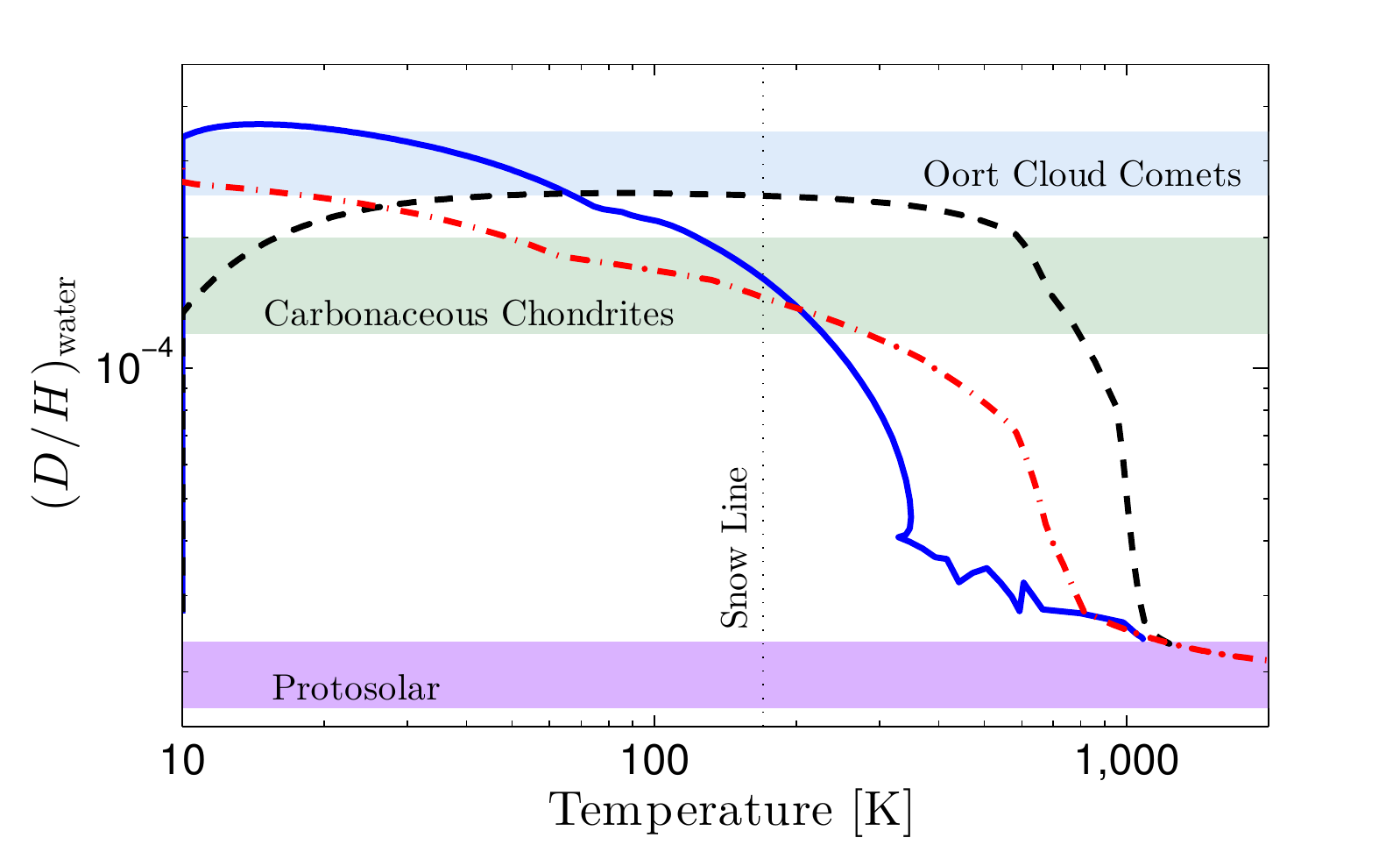}
\caption{The \dhw ratio shown against temperature after 1~Myr of evolution. Compared to several solar-system constrains: protosolar \citep[e.g.][]{GeissGloeckler2003}, carbonaceous chondrites \citep[e.g.][]{Alexanderetal2012} and the majority of Oort cloud comets \citep[e.g][]{BM09,BM12}. The snow-line at a temperature of $170$~K is marked with the thin vertical dotted line. The thick solid line shows a simulation with $Sc=1.0$, accretion bursts and $\Omega_c=8.88\times10^{-15}$~s$^{-1}$. The thick dashed line shows a simulation with $Sc=1.0$, no accretion bursts and $\Omega_c=7\times10^{-15}$~s$^{-1}$. The thick dot-dashed line corresponds to the best fit solution of \citet{jacquet13} with $Sc=0.16$.}\label{fig:DH_temp}
\end{figure}

  The importance of fact (ii) may be grasped from examination of Figure ~\ref{fig:DH_temp} which plots \dhw against the disc temperature for various disc models, along with the aforementioned solar system data. The dashed line shows the curve for a simulation without outbursts {$\Omega_c=7\times10^{-15}$~s$^{-1}$} and a Schmidt number of 1, for which the maximum D/H ratio roughly matches that displayed by comets at large radii ($\gtrsim 10$~AU). It is seen that at the snow line (taken here to correspond to a temperature of 170 K), \dhw is not very different from this maximum value, that is, in this case diffusion of equilibrated water from the inner solar system is too inefficient to induce a substantial deviation. However, carbonaceous chondrites, which generally show ample evidence of aqueous alteration and therefore initial presence of ice, should have formed beyond the snow line, and yet their D/H ratio is much lower than this. This \citet{jacquet13} pointed out as a general problem of standard disc models with $Sc\gtrsim 1$.  However, \citet{jacquet13} noted that assuming a Schmidt number of about 0.2 (more efficient turbulent diffusion), would alleviate this discrepancy, as illustrated by their analytic profile (in a quasi-static approximation) plotted as a red dot-dashed line in the figure\footnote{Note: \citet{jacquet13} adopt a different opacity and viscosity structure to our calculations. We have thus recalculated the equilibration temperature given by Jacquet \& Robert using our opacity and viscosity choices and obtain 777~K instead of 500~K. The best fit Schmidt number is however only slightly changed from 0.20 to 0.16 as the reduction of the equilibration radius is somewhat compensated by the lower equilibrium \dhw (at higher temperatures) there.}. While non-MRI and non-ideal turbulence may produce Sc values $\lesssim 1$ \citep[e.g.][]{Prinn1990,zhu14}, it is unclear how realistic such low Schmidt numbers are, and the required Sc values could be yet lower if the D/H values displayed by carbonaceous chondrite clays were compromised by isotopic exchange with organic matter as suggested by \citet{Alexanderetal2012}. If we now allow for bursts, the situation changes: the solid blue line in Figure ~\ref{fig:DH_temp} stands for a simulation with bursts ($\Omega_c=8.88\times 10^{-15}$~s$^{-1}$) also matching the cometary \dhw value at large radii, but this time matching the chondritic value at the snow line, despite a Schmidt number of 1. This is because accretion bursts have been able to process water in the inner disc, affecting the snow line, but the \dhw value at large radii has not evolved in pace with it (see also Figure \ref{fig:pre_post}). So this may be an alternative way to account for the low D/H ratios of carbonaceous chondrites.
%\begin{comment}
%In Figure~\ref{fig:DH_temp} we plot the \dhw ratio from several models compared to the disc temperature, where we over-plot various solar-system measurements of the \dhw ratio. We find two simulations that match the \dhw ratio in comets ($\sim 3\times10^{-4}$) at large radii ($\gtrsim 10$~AU), where both simulations have a rotation rate of $5-10\times10^{-15}$~s$^{-1}$. The simulations with no accretion burst and a Schmidt number of 1.0  (dashed line) matches the cometary value at large radius; however, the profile appears inconsistent with the value measured in carbonaceous chondrites, in the sense that diffusion from the inner solar system is too inefficient to make the snow line value deviate significantly from the cometary one. However, \citet{jacquet13} showed that a Schmidt number of 0.2 satisfied cometary and carbonaceous chondritic constraints for a burst-free disk model in the steady-state approximation, whose profile is plotted as a dot-dashed line in the figure\footnote{None of our burst-free simulations with Sc$=0.1$ matched the cometary value to be plotted for comparison to the analytical result of \citet{jacquet13}, for our chosen primordial D/H ratio of $10^{-3}$.}. Finally, the simulation with accretion bursts and a Schmidt number of 1.0 (solid line) also appears to agree well with the solar system constraints matching both the cometary and chondrite \dhw ratio at the expected radii of accretion.  
%\end{comment}

 To summarize, in order to explain the distribution of \dhw ratios measured in carbonaceous chondrites and comets we need that either: (i) the turbulent transport occurs with a rather low Schmidt number $\sim 0.2$, and the disc did not experience accretion bursts \citep{jacquet13}; or (ii) turbulent transport occurs with a nominal Schmidt number of $\sim 1$ and the disc experiences several accretion bursts during the initial stages of its evolution. 

\subsection{Caveats and avenues for further work}

%-Caveats (1d model - Yang ran a lot of other calculations so appeal to those), very simple chemical model
While we have shown that accretion bursts can have a significant impact on the evolution of the \dhw ratio, our results cannot be considered precise predictions. Rather, our simple models have allowed us to explore a large parameter space. Thus, we must discuss the impact our simplifications may have had on the evolution. In particular, we adopted a rather simple model for the chemical evolution of the water and integrated one rate equation instead of a large chemical network; however, we have shown our method agrees well with the more detailed calculations of \citet{yang13}, although reactions not considered by the latter (e.g. ion-neutral) could be important \citep{Albertssonetal2014}. 

Perhaps the biggest simplification is the use of a one-dimensional, one-layer model, where the evolution is controlled by the properties at the mid-plane of the disc. \citet{yang13} performed a series of simulations where they considered the impact of the vertical temperature structure on the evolution of the \dhw ratio and found it to be negligible. However, this may not be true for the layered structure in the dead-zone where the active surface layers that contain little mass, have a temperature different from the more massive dead zone \citep{%zhu11,
bae12,martin14} and there may be turbulent mixing between the active and dead layers, with the possibility that the dead zone itself contains residual levels of turbulence \citep[e.g.][]{FlemingStone2003,Turneretal2010,OkuzumiHirose2011}. Such a 2D structure would be worth taking into account.

%In addition to including a more detailed chemical network into the calculation \citep{yang13,Albertssonetal2014} one important next step would thus be the use of a two-layer model \citep[e.g.]{zhu10b,bae12}, where the active and dead zone can have different temperatures and water concentrations. Furthermore, it is expected that the dead zone may not be entirely dead and may contain residual levels of turbulence \citep[e.g.][]{FlemingStone2003,Turneretal2010,OkuzumiHirose2011} that could lead to radial and vertical mixing of water into and out of the dead-zone.

  We have also assumed throughout the paper that water was tightly coupled to the gas. However, this may not be true at large distances (where water is solid) and late stages of the disc's evolution, especially if water becomes incorporated in growing aggregates of mm to m size \citep[e.g.][]{CieslaCuzzi2006}. It might be \textit{a priori} expected that the effect will be to make outward transport of equilibrated water more difficult \citep{jacquet13} but simulations including gas-solid drift, accretion and fragmentation would be worthwhile to address the effect on the \dhw ratio.

Finally, since we have shown (in agreement with \citealt{yang13}) that the radius and time at which infalling material reaches the disc plays an important role in the late time \dhw profile, then more realistic disc formation models should be considered that account for differential rotation and turbulence in the cloud core, as well as the two- (and possibly three-) dimensional nature of the infall onto the disc \citep{visser09,visser11}, along with the associated angular momentum mis-match that can occur between the Keplerian disc rotation and the rotation rate of the infalling material. 
%-Further avenues of interest: not fully dead, dead-zones. Include bursts in better chemical models. More realistic star-formation models as disc building is important. 

\section{Conclusions}\label{sec:conclusion}

We have used a 1D model of a layered protoplanetary disc that undergoes accretion bursts due to the gravo-magnetic limit cycle (a commonly invoked mechanism to explain FU Orionis outbursts) to study their impact on the \dhw ratio. We have performed two types of calculations; firstly, a set of calculations with a steady infall rate onto the disc where we vary the infall rate, infall radius, Schmidt number, Prandtl number and whether or not the disc contains a dead zone. Secondly, we perform a series of calculation where we build the disc and star from a uniformly rotating isothermal cloud core, where we vary the rotation rate, Schmidt number and whether the disc contains a dead zone. In all these calculations we find that accretion burst play a major role in equilibrating water with the surrounding hydrogen gas. Thus, the \dhw ratio carries an imprint of whether the disc experienced accretion bursts or not. Our main findings are as follows:
\begin{enumerate}
\item During an accretion burst, water in the dead zone region is equilibrated, lowering its D/H ratio to $\sim 2\times10^{-5}$, as the dead zone temperature reaches $\sim 1000$~K during the accretion burst, higher than the equilibration temperature.
\item Discs that undergo accretion bursts have, in general, lower \dhw ratios as the burst thermally process a larger fraction of the disc material. Equilibrated water from the accretion burst can be carried to large radius by diffusion resulting in lower D/H ratios at 10-100s of AU.
\item Discs that undergo accretion bursts have larger equilibration radii (1-5~AU) compared to discs that do not experience accretion bursts. 
\item The rotation rate of the parent cloud core plays an important role in setting the large scale \dhw profile as suggested by \citet{yang13}, indicating there should be a significant variation in the D/H ratio in protoplanetary discs. 
\item Agreement with the solar-system constraints requires, either: (1) the solar nebula experienced no accretion bursts and had a low Schmidt number $\lesssim 0.2$; or (2) the solar nebula experienced accretion bursts and had a Schmidt number close to ``nominal'' (i.e. unity). 
\end{enumerate}

Given the important role of accretion bursts in the evolution of (D/H)$_{\rm water}$, we suggest more detailed modelling of their impact on the chemical and physical evolution of protoplanetary discs be performed. Finally, coupling these models with future observations, the \dhw ratio and profile could be used as probe for the prevalence of accretion bursts in the earliest stages of protoplanetary disc evolution.  

\section*{Acknowledgements}
We thank the referee for comments which improved the paper. We are grateful to Phil Armitage for useful discussions. The numerical calculations were performed on the Sunnyvale cluster at CITA, which is funded by the Canada Foundation for Innovation.
 
\appendix
\section{Numerical Test}
\label{sec:Owen vs Yang}
\begin{figure*}
\centering
\includegraphics[width=\textwidth]{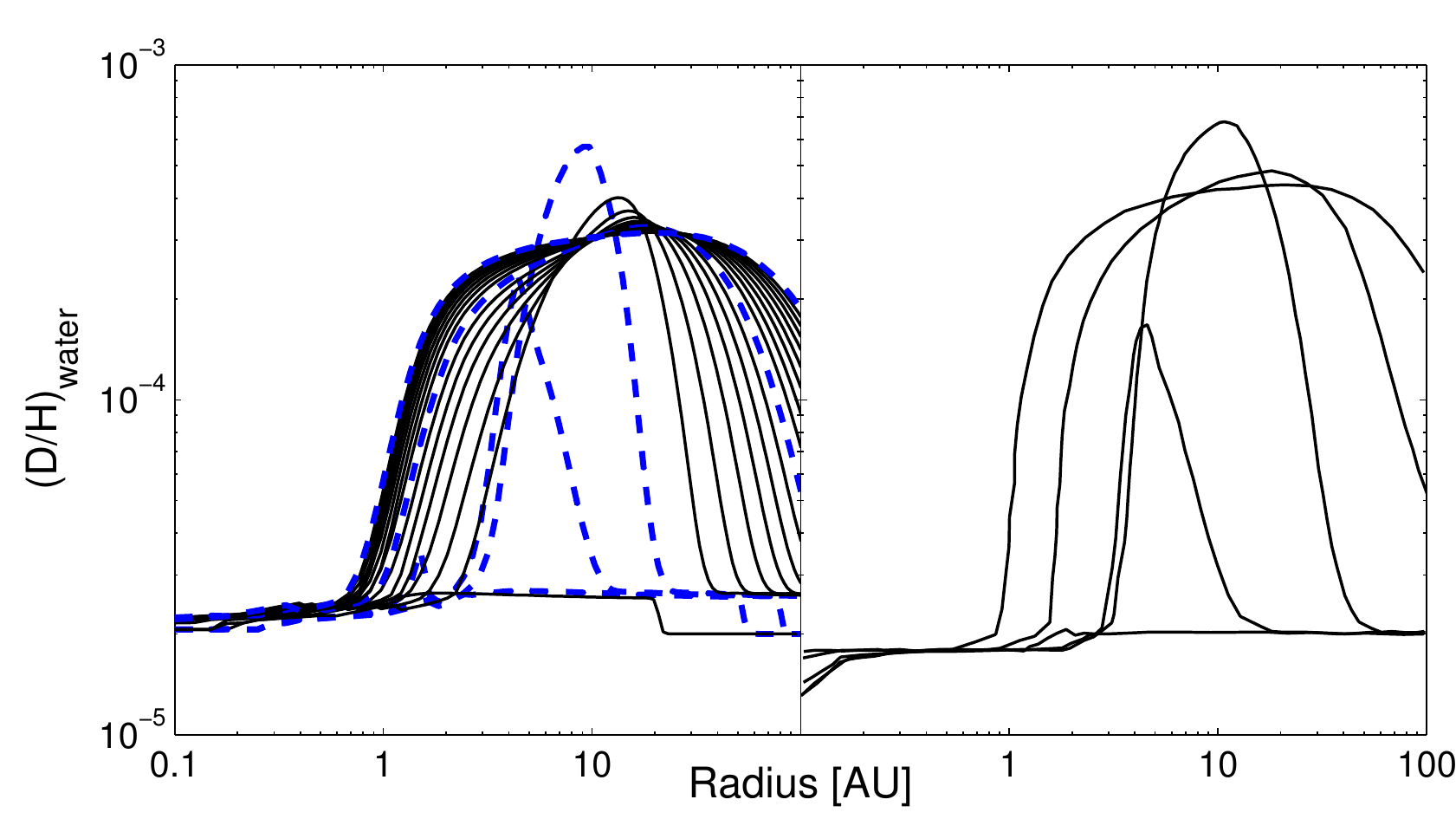}
\caption{Results of the benchmarking calculation. The left-hand panel shows the \dhw ratio plotted every 0.05 Myr from our calculation. The right-hand panel shows the results from the \citet{yang13} calculation (their `standard model' shown in Figure~4 of \citealt{yang13}) with the \dhw ratio plotted at 0.1, 0.2, 0.3, 0.5 \& 1.0 Myr. To ease comparison, we indicate the same times shown for the Yang et al. calculation in our calculation by plotting those times as thick dashed lines, rather than solid lines.}\label{fig:yang_bench}
\end{figure*}

Since we have adopted a simplified approach to the chemical evolution in view of the numerous chemical reactions that may govern evolution of the deuterium fraction in water, we need to test whether our method is appropriate for studying the effect of accretion bursts. Therefore, we repeat the `standard model' of disc evolution calculation of \citet{yang13} who include a much more complete chemical network with 14 species and 48 reactions. The model parameters are described in Table~3 of \citet{yang13} (i.e. $\Omega_c=10^{-14}$~s$^{-1}$, $T_c=15$~K \& $\alpha=0.001$). However, we do not simplify our temperature solver and opacity structure to that used by \citet{yang12,yang13}, but rather use the more detailed method described in Section~\ref{sec:1dmodel}. The results of our calculation are shown every 0.05Myr in the left hand panel of Figure~\ref{fig:yang_bench} and the results from the Yang et al.'s original calculation are shown in the right-hand panel at 0.1, 0.2, 0.3, 0.5 \& 1.0 Myr.

We see that the two evolutionary paths agree very well, particularly at late times (which we are primarily interested in from the point of view of observables). This indicates that our simple method for calculating the evolution \dhw ratio produces results comparable with the more detailed chemical method used by \citet{yang13}. This is not surprising since \citet{yang13} performed a detailed exploration of the different approaches for the chemical evolution and found they gave similar results. Therefore, we are confident that our method is suitable for exploring the effect of accretion bursts on the D/H ratio in water.

\end{document}